\documentclass[12pt]{iopart}
\begin{document}

\title[Wigner-function description of EPR experiment]{Wigner-function description of EPR experiment}

\author{Riccardo Franco
\footnote[3]{To whom correspondence should be addressed riccardo.franco@polito.it}
}
\address{Dipartimento di Fisica and U.d.R. I.N.F.M., Politecnico di Torino
C.so Duca degli Abruzzi 24, I-10129 Torino, Italia}

\date{\today}

\begin{abstract}
We provide a detailed description of the EPR paradox (in the Bohm
version) for a two qubit-state in the discrete Wigner function
formalism. We compare the probability distributions for two qubit
relevant to simultaneously-measurable observables (computed from
the Wigner function) with the probability distributions
representing two perfectly-correlated classic particles in a
discrete phase-space.
We write in both cases the updating formulae after a measure, thus
obtaining a mathematical definition of \textit{classic collapse}
and \textit{quantum collapse}.
We study, with the EPR experiment,  the joint probability
distributions of Alice's and Bob's qubit before and after the
measure, analyzing the non-local effects.
In particular, we give a more precise definition of locality,
which we call m-locality: we show that quantum systems may violate
this kind of locality, thus preserving, in an EPR-like argument,
the completeness of Quantum Mechanics.
%
%
%
%
\end{abstract}

\maketitle

%
%
%
%
\section{Introduction}
%
The Wigner function is a quasi-probability distribution which can
be used, in alternative to the density-matrix formalism, to
represent quantum states. The main advantage in the application of
Wigner function is that it behaves similarly to classical
probability distributions from several points of view. For quantum
states with infinite dimensional Hilbert-spaces, the Wigner
function has become a standard part of considerations. For finite
dimensional Hilbert-spaces, the Wigner-function formalism was
first investigated by Wootters \cite{Wootters} and later a more
geometric definition has been introduced \cite{Wootters_geom}. For
a list of references, see also \cite{Franco-Penna}. The discrete
Wigner function has shown to be useful in several applications:
the investigation of coherent states in a finite-dimensional basis
\cite{Coherent}, the definition of Q-functions and other
propensities \cite{Qfunction}, the development of number-phase
Wigner functions \cite{Phase-number}, the quantum tomography
\cite{Tomogr}, the quantum computation \cite{Galvao_speed}, the
theory of quantum games \cite{PazRoncagliaSaraceno}, the quantum
teleportation \cite{KBJ} and the study of entanglement
\cite{Franco-Penna}.

The EPR paradox \cite{EPR_art} plays a central role both in the
discussions about the interpretation of quantum mechanics and in
many quantum information applications. The original aim of the EPR
paradox was to entail the incompleteness of Quantum Mechanics.
More recently, EPR-type experiments, championed by Aspect et al.
\cite{Aspect}, are often interpreted as empirical evidence for the
existence of the "quantum non-locality". These non-local effects
are basic in all the main applications of Quantum Information
Theory.
The EPR paradox is a paradox in the following sense: if one takes
quantum mechanics and adds some seemingly reasonable conditions
(referred to as "locality", "realism", and "completeness"), then
one obtains a contradiction. However, quantum mechanics by itself
does not appear to be internally inconsistent, nor does it
contradict relativity.


The paper is organized as follows: in section
(\ref{Intro_collapse}) we give a brief review of some elements of
finite dimensional Wigner-function formalism (for a more detailed
review, see \cite{Wootters, Franco-Penna}) and we describe in
terms of Wigner function the measurement process. In section
(\ref{experiment}) the entire EPR paradox is discussed in terms of
discrete Wigner function, while in subsection (\ref{nonlocality})
the concept of non-locality is discussed and a more precise
definition is purposed. Finally in section (\ref{conclusions})
conclusions are drawn.

\section{The Wigner function and the measurement process}\label{Intro_collapse}
%
%
According to the definition given in \cite{Wootters}, the discrete
Wigner function relevant to a density matrix $\widehat{\rho}$ in a
$N=r^n$-dimensional Hilbert space $H$ (with $r$ prime number and
$n>0$) is a real function $W(\alpha)$, whose arguments $\alpha =
(\alpha_1, \alpha_2, .., \alpha_i, .. ,\alpha_n)$ represent the
coordinate in the whole phase space, each $\alpha_i=(q_i,p_i)$
being the coordinate relevant to the $i$-th subsystem. We can also
consider the coordinate $\alpha$ as a couple $(q,p)$ where each
coordinate is a multi-label, that is for example $q = (q_1, q_2,
... , q_i, ... , q_n)$. The explicit form of the Wigner function
is $W(\alpha)=\frac{1}{N}tr[\widehat{\rho} \widehat{A}(\alpha)]$,
where the operators $\widehat{A}(\alpha)$'s are the discrete
phase-point operators, forming a complete basis of the hermitian
operators acting on a $N=r^n$-dimensional Hilbert space. When
necessary, we will write the label $W_{\rho}(\alpha)$, evidencing
that the Wigner function is relevant to the density matrix
$\widehat{\rho}$, or $W_i(\alpha)$ where $i=a, b, ..$ labels
different steps. The general definition of $\widehat{A}(\alpha)$
is given in $\cite{Wootters}$, while we are interested in the
Wigner function relevant to a system of $n$ qubits. In the
particular case of $n=1$ (single qubit), the phase space is the
set of points $\alpha=(q,p)$, where $q, p\, = 0, 1$ are the
discrete phase-space coordinates (discrete position and discrete
momentum), while the discrete phase-point operators are written in
terms of Pauli matrices as \cite{Wootters} $
\widehat{A}(\alpha)=\frac{1}{2}\left[I+(-1)^{q}{\sigma}_z +
(-1)^{p} {\sigma}_x+(-1)^{q+p}{\sigma}_y\right] $.
Consistently with the definition of phase space in terms (of cartesian product) of constituent subspaces
\cite{Wootters}, phase-point operators $\widehat{A}({\alpha})$ are defined as tensor product of phase-point
operators relevant to the corresponding subsystems: $\widehat{A}({\alpha}) \, = \widehat{A}({\alpha_1}) \otimes
\widehat{A}({\alpha_2}) \otimes ... \otimes \widehat{A}({\alpha_n})$.
In the two qubit case, the Wigner function can be written
explicitly as $W(q_1, q_2, p_1, p_2)= \frac{1}{4}tr[\widehat{\rho}
\widehat{A}(q_1, p_1)\otimes \widehat{A}(q_2, p_2)]$.

The Wigner function has an important property, called the inner-product rule:
\begin{equation}\label{WFinner_discr}
tr(\widehat{\rho}
\widehat{\rho}')=N\sum_{\alpha}W_{\rho}(\alpha)W_{\rho'}(\alpha)\,.
\end{equation}
%

The phase-space operators $\widehat{Q}$ and $ \widehat{P}$,
relevant to the discrete phase-space coordinates $q$ and $p$, are
defined in terms of the sets of states $\{ \left|q\right> \}$ and
$\{ \left|p\right> \}$ (forming two orthonormal basis of $H$ and
connected by
$\left|p\right>=\frac{1}{\sqrt{N}}\sum_{q=0}^{N-1}e^{i\frac{2\pi}{N}qp}\left|q\right>$)
as \cite{KBJ}
\begin{equation}\label{axis_op}
\widehat{Q}=\sum_{0}^{N-1}q \left|q\right> \left<q|\right> \,\,\,
, \,\,\, \widehat{P}=\sum_{0}^{N-1}p
\left|p\right>\left<p\right|\, .
\end{equation}
%
%
%
%
%
%

Since the EPR experiment involves a measure on an entangled state,
we describe the measurement process within the discrete Wigner
function picture.
%
%
In general, quantum systems exist in a superposition of basis
states, and evolve according to the time-dependent Schr\"{o}dinger
equation (a process included in all interpretations of Quantum
Mechanics). For example, we consider a quantum state
$\left|s\right>$ in a finite dimensional Hilbert space and an
observable $\widehat{X}$, with a discrete set of eigenvalues
$\{x\}$ and eigenvectors $\{\left|x\right>\}$. Suppose we measure
the observable $\widehat{X}$: we could get, as a measurement
result, the value $\widetilde{x}$ (with probability
$|\left<s|\widetilde{x}\right>|^2$). After the measurement
process, the final state is $\left|\widetilde{x}\right>$. The
wave-function collapse is represented by this "jump" to one of the
basis states. After the collapse, the system begins to evolve
again according to the Schr\"{o}dinger equation. In the Copenhagen
interpretation \cite{Copen}, the collapse is a real process by
which quantum systems evolve according to the laws of quantum
mechanics.
%
%
%
We first analyze the analogue of a collapse situation in the case
of a classic probability distribution, where the effect of a
measurement is evidenced by the following:
\\
\textbf{Property A}: let us consider a set of independent
observable quantities $X, Y, Z, ...$, with discrete values $\{x,
y, z, ... \}$; we suppose that there is a generic initial joint
probability distribution $P_{a}(x, y, z, ...)$, representing the
probability of measuring contemporarily the values $(x,y,z,...)$.
If the measurement of $X$ gives a particular result
$\widetilde{x}$, then the joint probability distribution becomes
(\textit{classic-collapse} updating formula)
\begin{equation}\label{C_collapse}
P_{b}(x, y, z, ...)=\delta(x,\widetilde{x}) P_{a}(y, z, ...
|\widetilde{x}) \,\, ,
\end{equation}
which evidences that, for $x \neq \widetilde{x}$, $P_{b}$ is null
(since we have measured $\widetilde{x}$), while for
$x=\widetilde{x}$ the final distribution is the conditional
probability
$P(y,z,...|\widetilde{x})=P_a(\widetilde{x},y,z,...)/P_a(\widetilde{x})$.
Note that $P_a(\widetilde{x})$ is a marginal distribution,
obtained by summing $P_a(\widetilde{x},y,z,...)$ over the
variables $(y,z,...)$ and $\delta$ is the Kr\"{o}neker symbol.
%
%
In the special case of a probability distribution of one variable
$P_a(x)$ with measurement result $\widetilde{x}$, the final
probability distribution is simply
$P_b(x)=\delta(x,\widetilde{x})$: after the measure, we know with
certainty that the observable $X$ has the value $\widetilde{x}$.
%
%
%
%
%

%
In the quantum context all the probabilities can be computed from
$W(\alpha)$.
We first recall an important property of Wigner function, which
can be derived from equation (\ref{WFinner_discr}), valid for the
two qubit case (but easily extendible for $n$ qubit):
\\
\textbf{Property B}: given a two qubit state $\widehat{\rho}$ and
the corresponding Wigner function $W(\alpha)$, from the couples of
commuting operators $(\widehat{X}_1,\widehat{Y}_2)$ (with
$\widehat{X},\widehat{Y}=\widehat{Q},\widehat{P}$) we can write
the probability
$P(x_1,y_2)=\left<x_1,y_2|\widehat{\rho}|x_1,y_2\right>$ of
measuring contemporarily the values $(x_1,y_2)$. By writing the
Wigner functions
$W_{\left|x_1,y_2\right>}=W_{\left|x_1\right>}W_{\left|y_2\right>}$
, we have that
\begin{eqnarray}
\label{W_P_connection}
W(q_1,p_1,q_2,p_2)=\sum_{x_1,y_1=0,1}P(x_1,y_2)W_{\left|x_1,y_2\right>}(q_1,p_1,q_2,p_2)\\
\label{P_W_connection} P(x_1,y_2)=2 \sum_{\alpha}
W_{\left|x_1,y_2\right>}(q_1,p_1,q_2,p_2)W(q_1,p_1,q_2,p_2) \,\,.
\end{eqnarray}
Equation (\ref{W_P_connection}) can be used in the case of a
measure of observable $\widehat{X}_1$ with result
$\widetilde{x}_1$: the initial distribution $P_a$ is subjected to
a classic collapse, inducing on the initial state $W_a$ a
\textit{quantum collapse} thought the updating formula:
\begin{equation}\label{Q_collapse}
W_{b}(\alpha)=\sum_{x_1,y_1}P_b(x_1,y_2)W_{\left|x_1,y_2\right>}(\alpha)
=\sum_{x_1,y_1}\delta(x_1,\widetilde{x}_1)P_a(y_2|\widetilde{x}_1)W_{\left|x_1,y_2\right>}(\alpha)\,.
\end{equation}
It is important to note that the probability distributions
computed from (\ref{Q_collapse}) can violate the updating formula
(\ref{C_collapse}) of the classic case.
%
Equation (\ref{P_W_connection}) shows that in the case of the
observables $\widehat{Q}$ and $\widehat{P}$, defined by
(\ref{axis_op}), the related probabilities can be computed by
summing $W(\alpha)$ in the discrete phase-space over vertical and
horizontal lines respectively. In fact the Wigner functions
associated to states $\left|q\right>$ and $\left|p\right>$ have
graphics of vertical
and horizontal lines (see \cite{Wootters}).\\
We now compare the quantum and classic collapse in the simplest
example: given a single classic particle in a discrete phase space
with $q,p=0,1$, with a starting probability distribution
$P_a(q,p)$, if we measure $Q$ with value $\widetilde{q}$, by using
equation (\ref{C_collapse}) we have that the collapse produces
$P_b(q,p)=\delta(q , \widetilde{q})P_a(p\, | \, \widetilde{q})$.
In the corresponding quantum case, a single-qubit state, the
probability distribution (\ref{P_W_connection}) involves only
$\widehat{Q}$ or $\widehat{P}$ (since they are non-commuting
operators). If we measure $\widehat{Q}$ with value
$\widetilde{q}$, the starting probability distribution $P_a(q)$
becomes $P_b(q)=\delta(q,\widetilde{q})$ and thus, through
equation (\ref{Q_collapse}), the final Wigner function is
$W_b(q,p)=\frac{1}{2}\delta(q,\widetilde{q})$ (the Wigner function
associated to $\left|\widetilde{q}\right>)$. Thus
$P_b(p)=\frac{1}{2}$ (complete indetermination), different from
the classic result $P_b(p)=P_a(p|\widetilde{q})$.
%
%
\section{The EPR argument for a two-qubit system}\label{experiment}
%
The EPR argument \cite{EPR_art} can be applied both in the classic
and in the quantum case, by considering two non-interacting and
correlated systems in two spatially-separated laboratories. In the
first Alice can perform two different measures: we are interested
in which effects can have these measures on Bob's particle (second
laboratory).
In \cite{EPR_art} three hypothesis are considered as valid:
1) the \textit{reality}: if we can predict the value of an
observable without perturbing in any way the system, we have an
element of reality. For example, if we have the probability
distribution $P(x)=\delta(x,\widetilde{x})$ without performing any
measure on the system, then the observable $X$ is an element of
reality.
2) the \textit{locality}: given two distant and non-interacting
(in a certain time-interval) systems, the change in physical
properties of one system can not have direct influence on the
other in the same interval.
3) the \textit{completeness}: the description given by Quantum
Mechanics is complete.
%
%

First we consider the classic case, consisting in a two-particle
system with discrete observables $Q_i, P_j$ (with $i,j=1,2$) which
can be all measured contemporarily. The classic version of the EPR
state is a perfectly correlated two-particle state, whose
probability distribution  (with $q_i,p_j \in \{0,1\}$) is
\begin{equation}\label{CorrP}
P_{a}(q_1,p_1,q_2,p_2)=\frac{1}{4}\delta(q_1,q_2)\delta(p_1,p_2)
\, .
\end{equation}
In the EPR experiment, Alice can measure on the first particle
$Q_1$ or $P_1$: we suppose that Alice measures $Q_1$ with result
$\widetilde{q}_1$, which entails, from equation
(\ref{C_collapse}), the probability distribution
\begin{equation}\label{CorrPb}
P_{b}(q_1,p_1,q_2,p_2)=\frac{1}{2}\delta(q_1,\widetilde{q}_1)\delta(q_2,\widetilde{q}_1)\delta(p_1,p_2)
\,\, .
\end{equation}
The measure of Alice on her particle produces a change in the
probability distribution relevant to Bob's particle, as evidenced
by $P_b(q_2)=\delta(q_2, \widetilde{q}_1)$. Nevertheless, since
the two particles are spatially separated and non-interacting ,
from hypothesis 2) there is no physical change in Bob's system,
and the only change is in the knowledge of $Q_2$: thus $Q_2$ is an
element of reality. If Alice chooses to measure $P_1$ and obtains
$\widetilde{p}_1$, the probability distribution becomes
\begin{equation}\label{CorrPc}
P_{c}(q_1,p_1,q_2,p_2)=\frac{1}{2}\delta(p_1,\widetilde{p}_1)\delta(p_2,\widetilde{p}_1)\delta(q_1,q_2)
\,\, .
\end{equation}
Once again, the correlations of the particles allow us to entail
that $p_2=\widetilde{p}_1$, without measuring it and thus that
also $P_2$ is an element of reality.
From the two measures, we conclude that both $Q_2$ and $P_2$ are
elements of reality. Alice can also perform the two measurements
consecutively on her particle, and from equation
(\ref{C_collapse}) the distribution is
\begin{equation}\label{CorrPd}
P_{d}(q_1,p_1,q_2,p_2)=\delta(q_1,\widetilde{q}_1)\delta(q_2,\widetilde{q}_1)\delta(p_1,\widetilde{p}_1)\delta(p_2,\widetilde{p}_1)
\,\, .
\end{equation}
%

%
%
Let us now consider the quantum case: the Wigner function
corresponding to the EPR state $|\Psi \,+ \rangle = (|0,0\rangle
+|1,1\rangle)/\sqrt{2}$ (written in the base
$|0,0\rangle$,$|0,1\rangle$, $|1,0\rangle$, $|1,1\rangle$), which
results to be maximally entangled, is
\begin{eqnarray} \label{Wpsi}
W_{a}(q_1,p_1,q_2,p_2)= \frac{1}{8} - \frac{1}{4}\delta(q_1 \oplus
q_2, 1)\delta(p_1 \oplus p_2, 1) \\ \label{Wpsi1}
W_{a}^{T_2}(q_1,p_1,q_2,p_2)=\frac{1}{4}\delta(q_1,q_2)\delta(p_1,p_2)\,\,
,
\end{eqnarray}
where the symbol $\oplus$ denotes a sum mod2 and (\ref{Wpsi1}) is
the Wigner function of the partially transposed state, which is
identical to the classic distribution (\ref{CorrP}) (see
\cite{Wootters_geom} and \cite{KBJ}): functions (\ref{Wpsi}) and
(\ref{Wpsi1}) have the same marginal distributions.
From the initial state (\ref{Wpsi}) we can write the following
probability distributions
\begin{eqnarray}\label{P_x1x2}
P_a(q_1,q_2)=\frac{1}{2}\delta(q_1,q_2)\,,\,\,\,&
P_a(q_1,p_2)=\frac{1}{4}\\
\label{P_x1y2}P_a(p_1,p_2)=\frac{1}{2}\delta(p_1,p_2) \,,\,\,\,&
P_a(p_1,q_2)=\frac{1}{4}\,\,,
\end{eqnarray}
related to the couple of observables $(\widehat{Q}_1,
\widehat{Q}_2)$, $(\widehat{Q}_1, \widehat{P}_2)$ and
$(\widehat{P}_1, \widehat{P}_2)$, $(\widehat{P}_1,
\widehat{Q}_2)$, respectively.
If Alice measures $\widehat{Q}_1$ with value $\widetilde{q}_1$, we
have from equations (\ref{P_x1x2}) and (\ref{C_collapse}) that
$P_b(q_1,q_2)=\delta(q_1,\widetilde{q}_1)\delta(q_2,\widetilde{q}_1)$.
This is an effect of the correlations encoded in the EPR state:
without any measure, we perfectly know not only $\widehat{Q}_1$
but also $\widehat{Q}_2$. Since the second system is spatially
separated and non-interacting with the first, from the hypothesis
of locality we have that the second system can not be perturbed by
Alice's measurement: thus $\widehat{Q}_2$ is an element of
reality. The state resulting after the measure is
\begin{equation}\label{Wb}
W_{b}(q_1,q_2,p_1,p_2)=\frac{1}{4}\delta(q_1,\widetilde{q}_1)\delta(q_2,\widetilde{q}_1)
\,\, .
\end{equation}
The main difference from the classic case (\ref{CorrPb}) is that
function (\ref{Wb}) does not involve $\widehat{P}_1$ and
$\widehat{P}_2$, and thus the correlation between them has been
completely erased.
If Alice measures $\widehat{P}_1$ with result $\widetilde{p}_1$,
with similar arguments and by using equations (\ref{P_x1y2}), we
deduce that $\widehat{P}_2$ is an element of reality, while
$\widehat{Q}_2$ is undetermined after the measurement:
\begin{equation}\label{Wc}
W_{c}(q_1,q_2,p_1,p_2)=\frac{1}{4}\delta(p_1,\widetilde{p}_1)\delta(p_2,\widetilde{p}_1)
\, .
\end{equation}
Function (\ref{Wc}) does not involve $\widehat{Q}_1$ and
$\widehat{Q}_2$, and thus the correlation between them has been
completely erased.
The two measures performed by Alice allow to obtain two different
elements of reality, as written  in the distributions (\ref{Wb})
and (\ref{Wc}), relevant to two non-commuting operators. The EPR
conclusion is that, since hypotheses 1) and 2) should be
considered valid, the assumption 3) is wrong and the standard
description of Quantum Mechanics is incomplete.
In the EPR argument is implicit the idea that the \textit{context}
of the first measure does not influence the second: for example,
that the measure of $\widehat{Q}_1$ does not erase the correlation
between $\widehat{P}_1$ and $\widehat{P}_2$. This is true in the
classic case, where the two measures are independent, but not in
quantum mechanics, as can be seen from the updated functions
(\ref{Wb}) and (\ref{Wc}): thus the two measures are not
independent and the EPR conclusions must be considered in a
different way.
%
%
\subsection{The non-locality and the Wigner function}\label{nonlocality}
%
%
From the updated functions (\ref{Wb}) and (\ref{Wc}) we can deduce
that the measurements performed by Alice have influenced Bob's
qubit in a non-classic way, thus determining the differences in
the final results.
In the EPR argument the hypothesis of locality analyzes  the
change in physical properties of Bob's qubit as a direct influence
of Alice's measure. It can be easily shown that all the
probability distributions of Bob's qubit $P(q_2)$ and $P(p_2)$
change after Alice's measures following the classic updating
formula (\ref{C_collapse}), that is under the conditional
probability formula. This is sufficient to preserve the
no-communication theorem \cite{Peres-No comm} (fundamental for
special relativity) and to ensure no instantaneous information
transfer. On the contrary, we have evidenced that important
differences from the classic case arise in the quantum systems
when we consider joint probabilities of observables of the first
and second qubit.
We thus purpose a more precise definition of locality for a
probability distribution under a measurement process, which we
call \textit{measure-locality} or \textit{m-locality}:
\\
\textit{two distant and non-interacting systems (in a certain
time-interval) are measure-local (or m-local) when the measurement
of physical properties of the first system influences any
probability distribution of simultaneously-measurable observables
of the two systems, in the same time interval, according to the
updating formula }(\ref{C_collapse}).

In the classic case, if Alice measures ${Q}_1$  we have through
equation (\ref{CorrP}) that all the distributions $P(x_1,y_2)$
obey (\ref{C_collapse}), like for example
$P_a(p_1,p_2)=P_b(p_1,p_2)=\frac{1}{2}\delta(p_1,p_2)$. In the
quantum case, we have after measuring $\widehat{Q}_1$ and through
equations (\ref{P_x1x2}) and (\ref{Wb}),
\begin{equation}\label{m-non-locality}
P_a(p_1,p_2)=\frac{1}{2}\delta(p_1,p_2) \rightarrow
P_b(p_1,p_2)=\frac{1}{4}\,\,, ù
\end{equation} which violates the
classic updating formula (\ref{C_collapse}): the quantum system
shows a \textit{non-m-locality}. If we change, in the EPR
argument, the hypothesis 2) with the m-locality, we have that the
completeness of Quantum Mechanics is preserved.
Quantum systems are local (according to Einstein's definition),
but non-m-local.
%
%
\section{Conclusions}\label{conclusions}
%
%
%
We have described the EPR argument in terms of discrete Wigner
function: the use of a pseudo-probability function helped us to
directly compare the classic and the quantum situations.
%
%
We have written the explicit updating formulae (\ref{C_collapse})
and (\ref{Q_collapse}) for a measurement in the classic and
quantum case respectively, defining explicitly the concept of
classic collapse and quantum collapse in terms of probabilities.
%
%
We have evidenced through the Wigner function that the effect of a
measure on a quantum system is to erase certain correlations. This
has been evidenced by directly comparing the corresponding updated
functions (\ref{CorrPb},\ref{CorrPc}) and (\ref{Wb},\ref{Wc}) in
the EPR argument. We have concluded that, as an effect of the
quantum collapse (\ref{Q_collapse}), the two measures preformed by
Alice can not be performed independently and thus they are
contextual and somehow non-local.
%
%
We have pointed out that the marginal probabilities of Bob's qubit
change according to the classic updating formula, and thus do not
violate Einstein's locality principle. We have written a more
precise definition of locality, which refers to the joint
probability distributions of the two qubit, called
\textit{m-locality}. We have shown through equation
(\ref{m-non-locality}) that in the EPR argument the joint
probability distributions of Alice and Bob manifest a
non-m-locality.
%
%
We have purposed to modify the EPR argument by changing the
locality hypothesis with the m-locality, thus removing the
contradiction.\\
%
%
We notice that the anomalous change in the correlations between
subsystems before and after the measurement can be used to define
a new measure of entanglement, with a direct operational meaning:
partial results have been found on this respect, to be presented
in a separate paper. Other correlated considerations and results
are contained in \cite{Franco-Penna} (Local Uncertainty Relation).
\\\\
%
%
%

%
\end{document}